\documentclass[journal]{IEEEtran}
\ifCLASSINFOpdf
    \usepackage{graphicx}
\else
\fi
\usepackage{amsmath}

\DeclareMathOperator*{\argmin}{arg\,min}
\usepackage{amssymb}
\usepackage{dsfont}
\usepackage{bm}
\usepackage{xcolor}
\usepackage{algpseudocode} 
\usepackage{algorithm}
\ifCLASSOPTIONcompsoc
 \usepackage[caption=false,font=normalsize,labelfont=sf,textfont=sf]{subfig}
\else
 \usepackage[caption=false,font=footnotesize]{subfig}
\fi
\usepackage{placeins}

\begin{document}
\bstctlcite{IEEEexample:BSTcontrol} % This is to resolve the issue https://tex.stackexchange.com/questions/29381/is-it-normal-for-bibtex-to-replace-similar-author-names-with

% paper title
% Titles are generally capitalized except for words such as a, an, and, as,
% at, but, by, for, in, nor, of, on, or, the, to and up, which are usually
% not capitalized unless they are the first or last word of the title.
% Linebreaks \\ can be used within to get better formatting as desired.
% Do not put math or special symbols in the title.
\title{Interpretable Battery Cycle Life Range Prediction Using Early Cell Degradation Data}
%
%
% author names and IEEE memberships
% note positions of commas and nonbreaking spaces ( ~ ) LaTeX will not break
% a structure at a ~ so this keeps an author's name from being broken across
% two lines.
% use \thanks{} to gain access to the first footnote area
% a separate \thanks must be used for each paragraph as LaTeX2e's \thanks
% was not built to handle multiple paragraphs
%

\author{Huang~Zhang,
        Yang~Su,
        Faisal~Altaf, 
        Torsten~Wik,
        Sebastien~Gros
        % John~Doe,~\IEEEmembership{Fellow,~OSA,}
        % and~Jane~Doe,~\IEEEmembership{Life~Fellow,~IEEE}% <-this % stops a space
\thanks{Corresponding author: Huang Zhang is with the Department of Electromobility, Volvo Group Trucks Technology, 405 08 Gothenburg, Sweden (mobile: +46739022691; email: huang.zhang@volvo.com) and with the Department of Electrical Engineering, Chalmers University of Technology, 412 96 Gothenburg, Sweden (email: huangz@chalmers.se)}% <-this % stops a space
\thanks{Yang Su is with UMR ECOSYS, INRAE - UVSQ, Université Paris-Saclay, 78850 Thiverval-Grignon, France. (email: yang.su@inrae.fr)}
\thanks{Faisal Altaf is with the Department of Electromobility, Volvo Group Trucks Technology, 405 08 Gothenburg, Sweden (email: faisal.altaf@volvo.com)}% <-this % stops a space
\thanks{Torsten Wik is with the Department of Electrical Engineering, Chalmers University of Technology, 412 96 Gothenburg, Sweden (email: torsten.wik@chalmers.se)}
\thanks{Sebastien Gros is with the Department of Engineering Cybernetics, Faculty of Information Technology, NTNU, Gløshaugen, NO-7491 Trondheim, Norway (email: sebastien.gros@ntnu.no)}
% \thanks{Manuscript received April 19, 2021; revised August 26, 2015.}
}

\maketitle

% As a general rule, do not put math, special symbols or citations
% in the abstract or keywords. No more than 200 words
\begin{abstract}
Battery cycle life prediction using early degradation data has many potential applications throughout the battery product life cycle. For that reason, various data-driven methods have been proposed for point prediction of battery cycle life with minimum knowledge of the battery degradation mechanisms. However, managing the rapidly increasing amounts of batteries at end-of-life with lower economic and technical risk requires prediction of cycle life with quantified uncertainty, which is still lacking. The interpretability (i.e., the reason for high prediction accuracy) of these advanced data-driven methods is also worthy of investigation. Here, a Quantile Regression Forest (QRF) model, having the advantage of not assuming any specific distribution of cycle life, is introduced to make cycle life range prediction with uncertainty quantified as the width of the prediction interval, in addition to point predictions with high accuracy. The hyperparameters of the QRF model are optimized with a proposed alpha-logistic-weighted criterion so that the coverage probabilities associated with the prediction intervals are calibrated. The interpretability of the final QRF model is explored with two global model-agnostic methods, namely permutation importance and partial dependence plot. 

% The final QRF model not only provides cycle life point prediction with higher accuracy than an Elastic Net regression model, but also cycle life range prediction with high probability. Prediction uncertainties are quantified via prediction intervals, which enable both conservative and optimistic decisions for predicted battery cycle life, being important, for example, in the area of insurance and warranty.

% Revised 
% An accurate and calibrated of battery cycle life prediction using early degradation data is essential for most of applications throughout the battery product life cycle. Recently, massive data-driven methods have been proposed for point prediction of battery cycle life with minimum knowledge of the battery degradation mechanisms. However, they did not provide any information about prediction uncertainties, which is beneficial and profitable for the management of batteries at end-of-life with lower economic and technical risk. Inspired by quantile Regression Forest (QRF), in this work, a physics-informed QRF model designed for battery cycle life prediction is introduced, which provides not only point predictions with high accuracy but the calibrated quantified uncertainty. Furthermore, we proposed alpha-logistic-weighted criterion to calibrate the coverage probabilities associated with the prediction intervals.  

\end{abstract}

% Note that keywords are not normally used for peerreview papers.
% Up to 5 keywords

% This paper has not been presented at any conference or submitted elsewhere.

% \textcolor{blue}{Changes made in the resubmission of TTE-Reg-2022-03-0289 are colored in blue.}

% \textcolor{red}{Changes made in the revision of TTE-Reg-2022-07-1131 are colored in red.}

% For peer review papers, you can put extra information on the cover
% page as needed:
% \ifCLASSOPTIONpeerreview
% \begin{center} \bfseries EDICS Category: 3-BBND \end{center}
% \fi
%
% For peerreview papers, this IEEEtran command inserts a page break and
% creates the second title. It will be ignored for other modes.
\IEEEpeerreviewmaketitle

% According to "information for authors", Begin the main text of the paper on the second page.
% \newpage
\section{Introduction}
% The very first letter is a 2 line initial drop letter followed
% by the rest of the first word in caps.
% 
% form to use if the first word consists of a single letter:
% \IEEEPARstart{A}{demo} file is ....
% 
% form to use if you need the single drop letter followed by
% normal text (unknown if ever used by the IEEE):
% \IEEEPARstart{A}{}demo file is ....
% 
% Some journals put the first two words in caps:
% \IEEEPARstart{T}{his demo} file is ....
% 
% Here we have the typical use of a "T" for an initial drop letter
% and "HIS" in caps to complete the first word.

% You must have at least 2 lines in the paragraph with the drop letter
% (should never be an issue)
\IEEEPARstart{L}{ITHIUM-ION} (Li-ion) batteries have become the main choice of energy storage in electric vehicles (EVs), with a rapidly growing market that is spurred by governmental policies and subsidies with the aim of enhancing energy sustainability and carbon emission reduction \cite{hannan2017review} \cite{zubi2018lithium}. However, Li-ion batteries degrade with time due to both calendar aging and cyclic aging, which leads to a deterioration of their performance \cite{vetter2005ageing}. Understanding these aging processes, and providing a reliable cycle life prediction of Li-ion batteries based on early degradation data would enable many new possibilities throughout the battery life. We give five examples of such possibilities here. Firstly, the total driven distance of an EV battery can be translated into a number of equivalent full cycles \cite{zubi2018lithium}. Reliable cycle life prediction of Li-ion batteries using early degradation data would facilitate automotive companies to quickly adjust their warranty policy for new batches of Li-ion batteries from suppliers, while greatly reducing time and cost of long aging experiments. Secondly, warranty and pricing based on prediction of cycle life in a battery second life application as energy storage strongly affect how the second life battery market will evolve in the future, and reducing the uncertainty associated with cycle life prediction will reduce the cost of battery deployment \cite{martinez2018battery}. Thirdly, accurate and reliable cycle life prediction with high accuracy also facilitates predictive maintenance by reducing the sudden failure rate and the maintenance costs of battery-based applications \cite{severson2019data}. 
% Thirdly, prediction of cycle life with high accuracy using early degradation data can reduce the need of full aging characterization test for a large set of cells \cite{attia2020closed}. 
Fourthly, an early-prediction model can also be combined with a design parameter optimization algorithm to identify high-cycle-life charging protocols \cite{attia2020closed}. Lastly, accurate prediction of the battery life with early degradation data is of crucial importance for improving the battery development and manufacturing processes \cite{fermin2020identification}.

Unfortunately, accurate battery cycle life early prediction using relatively little degradation data that covers a limited range of lifetime is challenging, because the degradation process of Li-ion batteries is highly nonlinear with negligible capacity fade at early cycles, and influenced by not only the operating conditions, but also variances due to imperfect manufacturing tolerances. These factors contribute to the complexity of battery cycle life prediction \cite{baumhofer2014production}. This complexity and the importance of battery cycle life early prediction with high accuracy have made this an intense research area. Throughout the literature, the prediction methods can be generally divided into three categories - model-based methods, data-driven methods and hybrid methods.

The model-based methods, in turn, can be roughly classified into three categories. In the first one, a physics-based model, such as an electrochemical model (EM), is incorporated into a recursive filter framework, such as the extended Kalman filter \cite{huang2019aging} or a particle filter \cite{lyu2017lead}, in which internal parameters are updated from measured data. However, computational complexity in terms of high memory requirements and long computation time inevitably limits their applicability in real-time battery management system (BMS). In the second category, empirical models are identified based on cell characterization data from lab experiments. To improve their accuracy in on-board vehicle applications, model parameters can be adapted by the on-board BMS using measurement and state estimation data in a Bayesian filtering framework. This can include a range of Kalman filters \cite{chang2017new}, dual fractional-order extended Kalman filters \cite{hu2018co}, and particle filters \cite{zhang2018remaining}. The prediction accuracy of empirical models with a recursive filter highly depends on the fitted model. In addition to these two model-filter-based models, a third category of semi-empirical models have also been developed to capture the direct relationship between the operating conditions and the battery state-of-health (SoH), by interpolating and fitting experimental data. The only difference between empirical and semi-empirical model is that the former does not use any physical relation in the model structure whereas the latter uses some level of physical insights in the model formulation. Most semi-empirical models in the literature study the battery calendar aging and cyclic aging separately, and then combine both to make predictions under various operating conditions \cite{sarasketa2016realistic} \cite{schimpe2018comprehensive} \cite{de2017combined}. Although it is easier to implement semi-empirical models than the model-filter-based methods described, semi-empirical models are open-loop approaches where the model parameters are determined by data fitting.

Data-driven methods for battery cycle life prediction are generally black-box models developed based on machine learning or deep learning approaches to capture the mapping between inputs (e.g., features extracted from incremental capacity curves \cite{she2021battery}) and desired outputs (e.g., SoH). These methods can be either non-probabilistic or probabilistic. Non-probabilistic data-driven methods include autoregression (AR) based models \cite{long2013improved} \cite{zhou2016lithium}, artificial neural network (ANN) \cite{wu2016online} \cite{hu2015advanced} \cite{zhang2018long} and support vector machine (SVM) \cite{nuhic2013health} \cite{qin2015robust}. Despite the high accuracy of these non-probabilistic data-driven methods on cycle life point prediction, they are unfortunately not able to provide any uncertainty estimate of their predictions. The uncertainty level of predictions can enable a system or a user to make risk-informed decisions \cite{klas2018uncertainty}. For this reason, probabilistic data-driven methods like Gaussian process regression (GPR) \cite{richardson2019battery} \cite{liu2013prognostics} and relevance vector machine (RVM) \cite{wang2013prognostics} \cite{liu2015lithium} can be a better choice, as they have the ability to output a probability density function (PDF), and predict both the cycle life and the associated confidence interval. 
% However, the development of these probabilistic data-driven methods is still at an early stage. Few existing models are built on battery data under widely varying operating conditions, such as in an EV application, and the performance of these data-driven methods is highly dependent on the structure of the selected model and its hyperparameters. Therefore, their robustness and adaptation to real applications are uncertain.

Hybrid methods aim at leveraging the advantages of several different models. With the hybrid data-driven and model-based approach, a physics-based model is incorporated into a recursive filter framework (e.g., particle filter\cite{liao2016hybrid} \cite{wei2017remaining}, Kalman filter \cite{zheng2017remaining}), and the model parameters are identified and updated with measurements. Remaining useful life (RUL) is obtained by projecting estimated internal state to the future until a pre-defined end of life (EOL) threshold is reached. The data-driven model in this hybrid approach has been used for estimating the battery internal state from measurement data \cite{liao2016hybrid}, extrapolating the measurements beyond the range of currently available measurements \cite{liao2016hybrid} \cite{zheng2017remaining} and as a replacement of a degradation model \cite{wei2017remaining} in the physics-based model prediction case. The aforementioned studies have shown that hybrid methods have the potential to improve prediction accuracy further in comparison with one single data-driven model. However, they are difficult to use in online applications in a BMS due to its high computational complexity.

% The hybrid data-driven approach combines multiple data-driven models in order to improve the accuracy and reliability of the RUL prediction performance. There are two main approaches in this category. On the one hand, the prediction results of several data-driven models are aggregated to improve the prediction performance by a designed fusion mechanism (e.g., a weighted sum formulation \cite{hu2012ensemble}, a combination of the long-term signal and the regeneration signal \cite{liu2020data}). On the other hand, a data-driven model can be firstly used to estimate the internal state. The estimated internal state is then extrapolated to the future using another data-driven model \cite{gou2020state}.

% Establishing a niche
As illustrated by Severson et al. \cite{severson2019data}, cycle life for battery cells does not follow a normal distribution, which is a presumption of many probabilistic data-driven methods (e.g., GPR, RVM) that provide uncertainty information associated with the predictions. While advanced data-driven methods offer high prediction accuracy of battery cycle life in spite of minimum knowledge of the battery degradation mechanisms, interpretability of machine learning models is still underexplored in the literature. Extracting relevant battery aging knowledge from a machine learning or deep learning model in terms of underlying relationships, either in data or learned by the model, can provide valuable insights into battery aging. These insights can then be used to guide discoveries of aging mechanisms, improvements of battery manufacturing, and development of fast-charging protocols.

% Occupying the niche
This work tries to tackle the aforementioned problems by introducing a quantile regression forest (QRF) model for reliable cycle life range prediction of Li-ion battery cells. The prediction intervals (PIs) are constructed by using a quantile regression method that estimates quantiles of the response variable given values of the input variables \cite{koenker1978regression} \cite{koenker2005quantile}. The advantages of the QRF model over other probabilistic models are that asymmetric PIs can be estimated without assuming any specific distribution (e.g., Gaussian) of the output variable, i.e., the cycle life. Additionally, the QRF model is a non-parametric model, which means that the number of parameters automatically adapts to the complexity of the training data. 

The novelty and contributions are summarized as follows:

\begin{itemize}
\item 
This work proposes the first application of QRF model to provide battery cycle life point prediction and uncertainty range prediction. It is shown that the QRF model not only provides point prediction with high accuracy but also cycle life range prediction with high probability without assuming any specified distribution for the cycle life. The performance of the proposed QRF model is demonstrated on a public dataset that includes various operating conditions in terms of realistic charging current profiles. Its point prediction performance  is benchmarked to the Elastic Net model, whose exceptional early prediction performance was successfully demonstrated by Severson et al. \cite{severson2019data} and its range prediction performance is compared with two popular probabilistic models, i.e., GPR, RVM.

\item
An alpha-logistic-weighted criterion is proposed for optimizing hyperparameters of the QRF model and its effectiveness of improving the coverage probability of the final QRF model is demonstrated. The proposed criterion can also be used for optimizing hyperparameters of other regression models that are capable of providing range predictions.

\item
To statistically interpret the width of battery cycle life range prediction, two hypothesis tests are conducted. As a result, there is sufficient evidence in the first test that width of range prediction is highly correlated with absolute mean prediction error at a significance level of 0.05, which suggests that width of range prediction may provide more information for decision-making under uncertainties than we get from point predictions alone; there is also sufficient evidence in the second test that the width of range prediction is highly correlated with 6 input features at a significance level of 0.05, which suggests that the width of the range prediction is mainly affected by values of these 6 input features.

\item
To interpret the final QRF model for cycle life prediction, and reveal the underlying relationships in data learned by the QRF model, permutation importance and partial dependence plot are employed as model-agnostic methods to rank individual feature importance and quantitatively show the marginal effect each feature has on the predicted battery cycle life. Subsequently, an electrochemical interpretation is given to support what has been revealed by these two model-agnostic methods. 

\item
In an application of selecting the high-cycle-life charging protocol, the expected battery cycle life of a charging protocol can be determined with consideration of both point predictions and the uncertainty associated with the predictions. It is demonstrated that the final QRF model facilitates decision-making to select the high-cycle-life charging protocol that reduces the occurrence of unacceptably short cycle life.
% \item
% An area-based error score is proposed for tuning the hyperparameters of the QRF model so as to calibrate coverage probabilities associated with prediction intervals and later on for assessment of calibration quality on test set. Thereafter, an 85\% nominal coverage probability is selected to construct prediction intervals for battery cycle life range prediction on test set. Two pairs of cells with close point (e.g., mean) predictions (less than 5 cycles) are selected to show what more information is actually provided by the length of the prediction intervals.

% \item
% A sensitivity analysis of the QRF model performance using different number of early-cycle data (from the first 25 cycles to the first 250 cycles) shows that data for the first 100 cycles is a suitable choice for cycle life prediction, considering the trade-off between best possible prediction accuracy and lowest possible number of required cycles.

\end{itemize}

% \hfill mds
 
% \hfill August 26, 2015

\section{Theoretical background}

\subsection{Quantile Regression Forest}
% Start QRF
The random forest regression models approximate the conditional mean by a weighted sum over all the observations.
%with two steps, 1) calculate the conditional mean of each tree by simply averaging over all observations in the leaf node that contains the input; 2) summing all the predicted mean from each tree.
Instead of averaging over all observations in every leaf of every tree, one could use all observations from each tree to construct an empirical cumulative distribution function of the response variable.
Therefore, QRFs use full information of all the obsevartions via combining quantile methods and random forest. For simplicity, firstly, we denote the $\tau$-th quantile of $Y$ given $\bm{X}=\bm{x}$ by $q_{\tau}(Y | \bm{X}=\bm{x})$, where $\bm{X}$ is the input random variable, possibly high-dimensional, and $Y$ is the real-valued output random variable. The conditional cumulative distribution function $F(y|\bm{X}=\bm{x})$ is defined as the probability of $Y$ smaller than $y$ given $\bm{X}=\bm{x}$, i.e.,
\begin{equation}\label{eq: 6}
    F(y | \bm{X} = \bm{x}) = P(Y \leq y | \bm{X} = \bm{x}).
\end{equation}
For a continuous conditional cumulative distribution function $F(y|\bm{X}=\bm{x})$, as defined above, the $\tau$-th quantile $q_{\tau}(Y | \bm{X}=\bm{x})$ is defined such that the probability of $Y$ less than or equal to $q_{\tau}(Y | \bm{X}=\bm{x})$ is equal to $\tau$ for a given $\bm{X}=\bm{x}$, i.e.,
\begin{equation}\label{eq: 7}
    q_{\tau}(Y | \bm{X}=\bm{x}) = \inf \{ y : F(y|\bm{X} = \bm{x}) \geq \tau \}.
\end{equation}
% There are three hyperparameters that determine how a random forest is grown, i.e., the number of trees in the forest, the number of random features in each node split, and the minimum number of samples at a leaf node. 
During the inference, for an input $\bm{X}=\bm{x}$, the leaf of the $t$-th regression tree that contains $\bm{x}$ is denoted by $l(\bm{x}, \bm{\theta}_t)$, where $\bm{\theta}_t$ is the parameter vector that determines how the $t$-th tree is grown, for example, input variables that are considered in each node split. The weight $w_i(\bm{x} ,\bm{\theta}_t))$ from each tree is calculated by
\begin{equation} \label{eq: 2}
    w_i(\bm{x}, \bm{\theta}_t) = \frac{\mathds{1}_{\{\bm{x}_i \in {l(\bm{x}, \bm{\theta}_t)}\}}}{\#\{j:\bm{x}_j \in {l(\bm{x}, \bm{\theta}_t)} \}}, \, i=1,...,N, 
\end{equation}
where $N$ is the total number of the observations, and $\mathds{1}_{\{\bm{x}_i \in {l(\bm{x}, \bm{\theta}_t)}\}}$ is an indicator function equal to 1 if $\bm{x}_i \in {l(\bm{x}, \bm{\theta}_t)}$ and otherwise equal to 0. $\#\{j:\bm{x}_j \in {l(\bm{x}, \bm{\theta}_t)} \}$ denotes the total number of observations that are in the leaf $l(\bm{x}, \bm{\theta}_t)$.

Therefore, the weight from the whole random forest is defined as the average of $w_i(\bm{x}, \bm{\theta}_t)$ over all the $T$ regression trees grown, which reads as 
\begin{equation}\label{eq: 4}
    w_i(\bm{x}) = \frac{1}{T} \sum_{t=1}^{T} w_i(\bm{x}, \bm{\theta}_t).
\end{equation}

Furthermore, the constructed conditional cumulative distribution function of the QRF model is expressed as
\begin{equation}\label{eq: 8}
    F(y | \bm{X}=\bm{x}) = E(\mathds{1}_{\{Y \leq y\}} | \bm{X}=\bm{x}),
\end{equation}
where $\mathds{1}_{\{Y \leq y\}}$ is an indicator function and equal to 1 if $Y \leq y$ and otherwise equal to 0. $E(\mathds{1}_{\{Y \leq y\}} | \bm{X}=\bm{x})$ can be approximated by the weighted sum over the observations of $\mathds{1}_{\{Y \leq y\}}$. Thus, an empirical conditional probability function $\hat{F}$, given $\bm{X}=\bm{x}$, can be obtained as
\begin{equation}\label{eq: 9}
    \hat{F}(y | \bm{X}=\bm{x}) = \sum_{i=1}^N w_i(\bm{x}) \mathds{1}_{\{y_i \leq y\}},
\end{equation}
where the weights $w_i(\bm{x})$ are the same as defined in (\ref{eq: 4}) and the indicator function $\mathds{1}_{\{y_i \leq y\}}$ determines whether the weight will be counted or not, depending on the condition $y_i \leq y$.

Finally, the estimated $\tau$-th quantile $\hat{q}_{\tau}(Y | \bm{X}=\bm{x})$ is obtained by replacing $F(y | \bm{X} = \bm{x})$ in (\ref{eq: 7}) with $\hat{F}(y | \bm{X}=\bm{x})$ in (\ref{eq: 9}), i.e., 
\begin{equation} \label{eq: 10}
    \hat{q}_{\tau}(Y | \bm{X}=\bm{x}) = \inf \{ y : \hat{F}(y|\bm{X}= \bm{x}) \geq \tau \}.
\end{equation}

In addition, based on the standard random forest grown in the QRF model, the conditional mean of $Y$ given $\bm{X}=\bm{x}$ can also provide point predictions by,
\begin{equation}
    \hat{f}(\bm{x}) = \sum_{i=1}^N w_i(\bm{x})y_i,
\end{equation}
where the weights $w_i(\bm{x})$ are defined in (\ref{eq: 4}).

\subsection{Prediction Interval}
The PIs can be constructed from the conditional quantiles estimated by the QRF model. Specifically, the $(1-\alpha) \times 100\%$ PIs for output variable $Y$, given $\bm{X}=\bm{x}$, is constructed by 
\begin{equation} \label{eq: 11}
\hat{I}(\bm{x}) = [\hat{q}_{\alpha/2}(Y | \bm{X}=\bm{x}), \hat{q}_{1-\alpha/2}(Y | \bm{X}=\bm{x})].
\end{equation}
For example, the 95\% PI for the output $Y$ is estimated by $\hat{I}(\bm{x}) = [\hat{q}_{.025}(Y | \bm{X}=\bm{x}), \hat{q}_{.975}(Y | \bm{X}=\bm{x})]$,
which should be interpreted as; given $\bm{X}=\bm{x}$, a new observation of output $Y$ is in the interval $\hat{I}(\bm{x})$ with a probability of 95\%.

\subsection{Permutation Importance}
In order to understand the underlying battery degradation process, the goal of battery cycle life prediction should not only be limited to learn a regression function $\hat{f}$ that is capable of making battery cycle life predictions with high accuracy, but also to identify input variables from feature engineering that are the most important for the prediction accuracy of the learned model. A tool like variable importance can be helpful for identifying which input variables that are the most important and therefore should be measured with high precision \cite{fisher2019all}. 

% Decision trees can be visualized by a two-dimensional binary tree and therefore highly interpretable. However, linear combination of trees lose this important characteristic. Thus, the random forest regression model must be interpreted in a different way \cite{friedman2017elements}. 

The concept of variable importance was first introduced by Breiman \cite{breiman2001random} for random forests, in which variable importance of an input variable is measured by the decrease in predication accuracy when the values of this input variable are randomly permuted/shuffled in out-of-bag samples and then dropping the out-of-bag samples down the corresponding trees. Louppe et al. \cite{louppe2013understanding} characterized an alternative measure of variable importance based on the Mean Decrease Impurity (MDI). Impurity is quantified by the splitting criterion of the decision trees (e.g., mean squared error for continuous outputs). However, variable importance based on MDI method favors high cardinality input variables over low cardinality input variables, such as binary variables or categorical variables with a small number of possible categories. Furthermore, variable importance based on the MDI method can only be calculated on the training set during growth of trees. Therefore, there are possibilities that the MDI method  gives high importance to input variables that may not be predictive on unseen data when the model is overfitted \cite{louppe2013understanding}. With the aim of mitigating those limitations of variable importance based on the MDI method, permutation importance is developed as a generalized model-agnostic method for measuring the importance of an input variable by calculating the increase of prediction error after permuting values of the input variable \cite{fisher2019all}. An input variable is considered to be important if shuffling its value leads to an increase of the model error and vice versa. Permutation importance can be computed on both the training set and the test set, which makes it possible to identify features that contribute the most to the generalized prediction power of the fitted model. The permutation importance algorithm below describes how the measure is calculated.

\begin{algorithm}
	\caption{Permutation importance algorithm} \label{algorithm_1}
% 	\hspace*{\algorithmicindent} \textbf{Input: }Trained model f, feature matrix \bm{X}, target vector y, error score S(y,f)
	\begin{algorithmic}[1]
	    \State \textbf{Input:} Learned model $\hat{f}$, training or test set $\mathcal{G}$, and error score function $s$ (e.g., $R^2$ score)
	    \State Compute the reference error score $s_0$ of the learned model $\hat{f}$ on dataset $\mathcal{G}$
		\For {each input variable $j=1,2,\ldots,p$}
		    \For{each repetition $m=1,2,\ldots,M$}
		    \State Randomly shuffle the values in the column corresponding to input variable $j$ in the dataset $G$ to generate a corrupted version of the dataset, $\widetilde{\mathcal{G}}$
		    \State Compute the error score $s_{j,m}$ of the learned model $\hat{f}$ on corrupted version of the dataset $\widetilde{\mathcal{G}}$
		    \EndFor
			\State Compute importance $\mathrm{Im}_j$ for input variable $j$ defined as $\mathrm{Im}_j = s_0 - \frac{1}{M} \sum_{m=1}^{M} s_{j,m}$
		\EndFor
	    \State Sort input variables in descending order of variable importance.	
	\end{algorithmic} 
\end{algorithm}

% Permutation importance provides condensed, global insights of the learned model $\hat{f}$ by randomly shuffling value of the feature, both the main feature effect on response variable and the interaction effects with other features on model performance are considered.

\subsection{Partial Dependence Plot}
After the most important input variables have been identified, the next step is to understand the dependence of the approximation $\hat{f}(\bm{X})$ on the joint values of the input variables \cite{friedman2001greedy}. 

Consider the subvector $\bm{X}_\mathcal{S}$ of length $\ell < p$ of the input vector $\bm{X} = (X_1, X_2, ..., X_p)^T$, indexed by $\mathcal{S} \subset \{1,2,...,p\}$. Let $\mathcal{C}$ be the complement set, with $\mathcal{S} \cup \mathcal{C} = \{1,2,...,p\}$, and $\bm{X}_\mathcal{C}$ the corresponding subvector. In principle, the approximation $\hat{f}(\bm{X})$ depends on all of the input variables, i.e., $\hat{f}(\bm{X}) = \hat{f}(\bm{X}_{\mathcal{S}}, \bm{X}_{\mathcal{C}})$. However, if the variables in $\bm{X}_{\mathcal{S}}$ do not have strong interactions with those in $\bm{X}_{\mathcal{C}}$, then the average, or partial dependence, of $\hat{f}(\bm{X})$ on $\bm{X}_{\mathcal{S}}$ is approximately
\begin{equation}
    \hat{f}_{\mathcal{S}}(\bm{X}_{\mathcal{S}}) = E_{\bm{X}_\mathcal{C}}[\hat{f}(\bm{X})] = E_{X_\mathcal{C}} [ \hat{f}(\bm{X}_\mathcal{C}, \bm{X}_\mathcal{S})] \label{pdp_1}.
\end{equation}
% where $\bm{x}_l$ is the chosen feature subset for which the PDP should be plotted, and $\bm{x}_{{\setminus} l}$ is the complement feature subset in the model $\hat{f}$. Here, $p_{{\setminus} l}(\bm{x}_{{\setminus} l})$ is the marginal probability density of $\bm{x}_{{\setminus} l}$. By marginalizing the model output over the distribution of the features in the set $\bm{x}_{{\setminus} l}$, we get a function $\bar{f}_{l}(\bm{x}_l)$ which describes the partial dependence between the features in set $\bm{x}_l$ and the predicted response variable. The marginal probability density of $\bm{x}_{{\setminus} l}$ is
% \begin{equation}
%     p_{{\setminus} l} (\bm{x}_{{\setminus} l}) = \int p(\bm{x}) d \bm{x}_{{\setminus} l} \label{pdp_2}
% \end{equation}
% where $p(\bm{x})$ is the joint density of all of the input features $\bm{x}$. The marginal probability density (\ref{pdp_2}) can be estimated from the training set, so that equation (\ref{pdp_1}) becomes
In practice, the partial dependence function $\hat{f}_{\mathcal{S}}(\bm{X}_{\mathcal{S}})$ can therefore be estimated by
\begin{equation}
    \bar{f}_{\mathcal{S}}(\bm{X}_{\mathcal{S}}) = \frac{1}{N} \sum_{i=1}^{N} \hat{f}(\bm{X}_{\mathcal{S}}, \bm{x}_{i \mathcal{C}}),
\end{equation}
where $\{\bm{x}_{1 \mathcal{C}}, \bm{x}_{2 \mathcal{C}}, ..., \bm{x}_{N \mathcal{C}} \}$ are the values of $\bm{X}_{\mathcal{C}}$ occurring in the training set, and $N$ is the total number of samples in the training set.

\subsection{Performance Evaluation Metrics}
\subsubsection{Evaluating Point Prediction Quality}
There are various performance evaluation metrics that can be used to evaluate the accuracy of point predictions. The most common ones are root mean squared error (RMSE), mean absolute percentage errors (MAPE) and coefficient of determination ($R^2$), defined as follows:
\begin{equation}
    \mathrm{RMSE}(y_i, \hat{y}_i) = \sqrt{\frac{1}{N_T} \sum_{i=1}^{N_T} (y_i - \hat{y}_i)^2}
\end{equation}
\begin{equation}
    \mathrm{MAPE}(y_i, \hat{y}_i) = \frac{1}{N_T} \sum_{i=1}^{N_T} \left | \frac{y_i - \hat{y}_i}{y_i} \right |
\end{equation}
\begin{equation}
    R^2(y_i, \hat{y}_i) = 1- \frac{\sum_{i=1}^{N_T} (y_i - \hat{y}_i)^2}{\sum_{i=1}^{N_T} (y_i - \bar{y})^2}
\end{equation}
where $N_T$ is the number of samples to be evaluated (i.e., all samples in the test set), $\hat{y}_i = \hat{f}(\bm{x}_i)$ denotes the estimated mean cycle life predicted by the model, $y_i$ denotes the corresponding observed cycle life, and $\bar{y} = \frac{1}{N_T} \sum_{i=1}^{N_T} y_i$ is the average cycle life for a total of $N_T$ samples in the test set.

\subsubsection{Evaluating Range Prediction Quality}
Two commonly used metrics for range predictions are prediction interval coverage probability (PICP) and mean prediction interval width (MPIW) \cite{khosravi2010construction}.

The PICP shows the percentage of output values covered between the lower and upper bounds of the PIs and as such assesses the calibration of the range predictions. A larger PICP means that more output values will fall in the constructed PIs, and vice versa. The PICP is defined as,
\begin{equation}
    \mathrm{PICP} = \frac{1}{N_T} \sum_{i=1}^{N_T} c_i \times 100\%,
\end{equation}
where $c_i=1$, if the output value $y_i$ is covered by the interval from the lower bound $L_i$ to the upper bound $U_i$ of the constructed PI; otherwise $c_i=0$.

In a practical application, such as the battery cycle life prediction in the present work, the width of constructed PIs is of equal importance as the coverage probability since it does not make much sense to have PIs with high coverage probability and large width at the same time. For a narrower PI, the prediction is clearly more informative than for a wider PI. Therefore, there is also a need to assess the sharpness of range predictions, which can be done via MPIW, defined as,
\begin{equation}
    \mathrm{MPIW} = \frac{1}{N_T} \sum_{i=1}^{N_T} | U_i - L_i |.
\end{equation}
Theoretically, it is desirable to have PIs with a PICP value close to their nominal coverage (e.g., 95\%) and a small MPIW value.

Both PICP and MPIW only assess PIs from one aspect. A comprehensive measure of both coverage probability and width of PIs at the same time is the averaged interval score (AIS), proposed in Ref. \cite{gneiting2007strictly}, which is defined as
\begin{equation} \label{eq: 12}
\begin{aligned}
    \mathrm{AIS} = & \frac{1}{N_T} \sum_{i=1}^{N_T} ((U_i-L_i) + \frac{2}{\alpha} (L_i-y_i) \mathds{1}_{\{y_i<L_i\}} \\ 
                   & + \frac{2}{\alpha} (y_i-U_i) \mathds{1}_{\{y_i>U_i\}}).    
\end{aligned}
\end{equation}
The AIS defined above rewards narrow PIs and penalizes intervals missed by the observation with a weight depending on $\alpha$, as defined in Eqn. (\ref{eq: 11}). The AIS is used for standard QRF training.

\section{Methodology and problem formulation}
\subsection{Feature Engineering and Selection}
Generally, feature engineering can be divided into two categories - manual feature engineering based on domain knowledge \cite{guo2019data} \cite{williard2013comparative} and automatic feature engineering, such as auto-encoders \cite{yousefi2017autoencoder}, and restricted Boltzmann machine \cite{bi2019enhanced}.

In this work, manual feature engineering based on battery domain knowledge is adopted. More specifically, 33 features were extracted from data of the first 100 cycles, at which point most batteries have not yet exhibited any significant capacity degradation. The 33 features are divided into five groups and listed in Table. \ref{tab01}.
\begin{table*}[!htbp]
\caption{33 features in 5 groups}
\begin{center}
\begin{tabular}{|c|l|}
\hline
\textbf{Groups} & \textbf{Features} \\
\hline
Charge-related & \shortstack[l]{Average charge time for the first 5 cycles \cite{severson2019data}.}\\
\hline
Discharge voltage curve-related & \shortstack[l]{Minimum, variance, skewness, and kurtosis of difference of the discharge voltage curve \\ between cycle 100 and cycle 10 (i.e., $\Delta Q_{100-10} (V)$) \cite{severson2019data}. \\ Amplitude and position shift of the highest peak in the discharge incremental capacity curve \\ between cycle 10 and cycle 100 (i.e., $dQdV_{100-10}$).} \\
\hline
Capacity-related & \shortstack[l]{Slope of the linear fit to the capacity fade curve from cycle 2 to cycle 100 \cite{severson2019data}. \\ Intercept of the linear fit to capacity fade curve from cycle 2 to cycle 100 \cite{severson2019data}. \\ Discharge capacity at cycle 2 \cite{severson2019data}.\\ Discharge capacity at cycle 100 \cite{severson2019data}.\\ Difference between maximum discharge capacity within the first 100 cycles and discharge \\capacity at cycle 2 \cite{severson2019data}.}\\
\hline
Temperature-related & \shortstack[l]{Minimum, variance, skewness, and kurtosis of difference in the discharge cell temperature, \\as a function of voltage, between cycle 100 and cycle 10 (i.e., $\Delta T_{100-10}(V)$). \\ Minimum, maximum, mean, and variance of discharge cell temperature as a function of \\voltage at cycle 10 (i.e., $T_{10} (V)$). \\ Minimum, maximum, mean, and variance of discharge cell temperature as a function of \\voltage at cycle 100 (i.e., $T_{100} (V)$). \\ Difference in minimum, maximum, mean, and variance of discharge cell temperature, \\as a function of voltage, between cycle 10 and cycle 100.}\\
\hline
Internal resistance-related & \shortstack[l]{Minimum internal resistance from cycle 2 to cycle 100 \cite{severson2019data}. \\ Maximum internal resistance from cycle 2 to cycle 100. \\ Internal resistance at cycle 2 \cite{severson2019data}. \\ Internal resistance at cycle 100. \\ Difference in internal resistance between cycle 100 and cycle 2 \cite{severson2019data}.}\\
\hline
% \multicolumn{6}{l}{Sample of a Table footnote.}
\end{tabular}
\label{tab01}
\end{center}
\end{table*}

To reduce the computation time, and improve the performance of learned models, a feature selection method is employed which selects an effective subset of features by reducing irrelevant and redundant features \cite{chandrashekar2014survey}. For this, a random-forest-based recursive feature elimination with cross-validation (RF-RFE-CV) is employed, which selects a subset of features by recursively removing features with the least importance in the current feature set. Only the training set is used for feature selection to avoid introducing optimistically biased performance estimates. As a result, 12 features were automatically selected as a feature subset and then fed into the QRF and Elastic Net models for battery cycle life prediction.
% The feature selection aims to select an effective subset of features via reducing irrelevant and redundant features. 

\subsection{Problem  Formulation}
The battery cycle life early prediction problem can be formulated as a regression problem with the goal of learning a mapping $f$ from a random input vector (a term we will use interchangeably with features) $\bm{X} = (X_1, X_2, ..., X_p)^T$ in the space $\mathcal{X} \subseteq \mathbb{R}^{p}$ to a random output (a term we will use interchangeably with response) variable $Y$ in the space $\mathcal{Y} \subseteq \mathbb{R}^{+}$, $f:\mathcal{X} \to \mathcal{Y}$, given a training set $\mathcal{D} = \{\bm{x}_{i}, y_{i}\}_{i=1}^N$, where $N$ is the number of assumed independent and identically distributed samples in the training set. In the present case, $\bm{x}_i \in \mathcal{X}$ represents $p$ features extracted from the first 100 cycles, and $y_i \in \mathcal{Y}$ is the observed battery cycle life.

To learn the mapping function $f$, the conditional mean minimizing the expected squared error loss, with the assumption that squared error loss function is symmetric around zero, is used,
\begin{equation} \label{eq: 1}
     E(Y | \bm{X} = \bm{x}) = \argmin_{f(\bm{x})} E\{(Y-f(\bm{x}))^2 | X = \bm{x}\}.
\end{equation}
In practice, the approximation of the conditional mean is achieved by minimization of a squared error type loss function over the training set $\mathcal{D}$, and the resulting learned regression function is denoted as $\hat{f}$.

% To indicate why there is a need for range prediction by quantile regression
The conditional mean only reveals one aspect of the conditional distribution of a response variable $Y$ and gives no information about the uncertainty associated with the predicted conditional mean. In our case, though, we are interested to find the range of predicted battery cycle life in which the battery will reach its end of life with high probability. We propose QRF to handle this case, yielding both a point prediction as well as its uncertainty. More specifically, its point prediction is provided by the estimated conditional mean, given $\bm{X}=\bm{x}$, i.e.,
\begin{equation}\label{eq: 20}
    \hat{f}(\bm{x}) = \sum_{i=1}^N w_i(\bm{x})y_i,
\end{equation}
where the weights $w_i(\bm{x})$ are defined in (\ref{eq: 4}). 

The corresponding range prediction, provided by a 95\% PI given $\bm{X}=\bm{x}$, is given by
\begin{equation}\label{eq: 11.1}
    \hat{I}(\bm{x}) = [\hat{q}_{.025}(Y | \bm{X}=\bm{x}), \hat{q}_{.975}(Y | \bm{X}=\bm{x})].
\end{equation}

\subsection{Proposed PI Evaluation Criterion}
% \subsection{Proposed training method}
From a decision-making perspective, having point predictions with a low RMSE and range predictions with a high PICP together with a low MPIW is preferable. 
% From the optimization perspective, the hyperparameter optimization problem can be formulated as a constrained three-objective problem as follows:
% \begin{subequations}
% \begin{align}
% &\text{Objectives}:  &\ &\text{Finding an optimal set of hyperparameters to,}\\
% &                    &\ &\text{maximize}: \mathrm{PICP}\\
% &                    &\ &\text{minimize}: \mathrm{RMSE}, \mathrm{MPIW}\\
% &\text{Constraints}: &\ & 0 \leq \mathrm{PICP} \leq 100\%\\
% &                    &\ & \mathrm{RMSE} > 0\\
% &                    &\ & \mathrm{MPIW} > 0
% \end{align}
% \end{subequations}
However, there is a trade-off between maximizing PICP and minimizing MPIW. The AIS defined in Eqn. (\ref{eq: 12}) only assesses PIs over each validation sample in the training set without consideration of whether a preassigned nominal coverage probability $(1-\alpha) \times 100\%$ is satisfied or not on the training set. Therefore, it is expected to propose a comprehensive measure that includes both two properties of PIs. More importantly, the calibration property is prioritized rather than the sharpness property of the PIs. Thus, higher penalties should be given in case of unsatisfactory nominal coverage probability $(1-\alpha) \times 100\%$ that is usually preassigned by a decision maker. To address such problems, we propose an alpha-logistic-weighted (ALW) criterion based on the work by Khosravi et al. \cite{khosravi2010construction}, which reads as 
\begin{equation}\label{eq: 21}
    \mathrm{ALW} = \frac{\mathrm{MPIW}}{\sigma(\alpha, \mathrm{PICP})},
\end{equation}
where $\sigma(\cdot)$ is the logistic function defined as,
\begin{equation}\label{eq: 22}
    \sigma(\alpha, \mathrm{PICP}) = \frac{1}{1+e^{-\frac{1}{\alpha}(\mathrm{PICP}-(1-\alpha))}},
\end{equation}
where $\alpha$ is the same as in Eqn. (\ref{eq: 11}). The terms $\frac{1}{\alpha}$ and $1-\alpha$ determine the growth rate and the midpoint of the logistic curve, respectively.
By inserting  Eqn. (\ref{eq: 22}) into Eqn. (\ref{eq: 21}), Eqn (\ref{eq: 21}) can be rewritten as
\begin{equation}\label{eq: 23}
    \mathrm{ALW} = \mathrm{MPIW}(1 + e^{-\frac{1}{\alpha}(\mathrm{PICP}-(1-\alpha))}).
\end{equation}
The ALW defined above rewards lower MPIW and exponentially penalizes unsatisfactory PICP that is lower than nominal coverage probability $(1-\alpha) \times 100\%$ with a weight depending on $\alpha$.

\subsection{Hyperparameter Optimization}
In this work, the objective of hyperparameter optimization is to find an optimal set of hyperparameters for the QRF model that minimizes the value of AIS, defined in Eqn. (\ref{eq: 12}) or the proposed ALW defined in Eqn. (\ref{eq: 23}), given a training set at each train-test split.  The leave-one-out cross validation (LOO-XVE) method \cite{friedman2017elements} is adopted for evaluating the averaged performance of a QRF model given a set of hyperparameters. The LOO-XVE is suitable for small datasets, as in the present work, where prediction performance also outweighs computational cost at the training stages. Optuna \cite{akiba2019optuna}, a Bayesian hyperparameter optimization framework, is used to search for the optimal set of hyperparameters for the QRF model given the training set. The final QRF model with the optimal set of hyperparameters is learned on the whole training set and is then evaluated on the test set at each train-test split.

\subsection{Proposed expected battery cycle life of a charging protocol}
In an application of selecting the high-cycle-life charging protocol, the expected battery cycle life (EBCL) of a charging protocol can be calculated by averaging over all predicted mean cycle lives of cells charged with this charging protocol \cite{attia2020closed}, 
\begin{equation} \label{eq: ecl1}
    \mathrm{EBCL} = \frac{1}{B} \sum_{i=1}^{B} \hat{f}(\bm{x}_i),
\end{equation}
where $B$ denotes the total number of battery cells charged with this charging protocol.

However, it will be a difficult selection decision to be made when the expected battery cycle lives for two charging protocols calculated using $\mathrm{EBCL}$ are very close to each other. Therefore, we propose the expected battery cycle life range (EBCLR) that helps facilitating decision-making in this case, 
\begin{equation} \label{eq: ecl2}
    \mathrm{EBCLR} = |U-L|,
\end{equation}
where $U$ and $L$ are the upper bound and lower bound respectively of the expected battery cycle life range prediction.
Considering the relatively small dataset used in this work, we adopt the median approach to combine PIs for each charging protocols \cite{grushka2020combining},
\begin{equation}
    L=\mathrm{Median}(L_1, ..., L_B); \ U = \mathrm{Median}(U_1, ..., U_B),
\end{equation}

where $L_1, ..., L_B$ are the lower bounds of battery cycle life range predictions of a group of $B$ cells that are charged with the same charging protocol, and $U_1, ..., U_B$ are the upper bounds of battery cycle life range predictions of a group of $B$ cells that are charged with the same charging protocol.

% both point prediction and range prediction are available, the expected battery cycle life of a charging protocol can be calculated as a weighted sum over all predicted mean cycle lives with the weight inversely proportional to the width of prediction intervals,
% \begin{equation} \label{eq: ecl2}
%     \mathrm{EBCL}_2 = \sum_{i=1}^{B} v_i \hat{f}(\bm{x}_i),
% \end{equation}
% where $v_i$ is the weight that is defined as
% \begin{equation} 
%     v_i = \frac{\frac{1}{U_i-L_i}}{\sum_{i=1}^{B} \frac{1}{U_i-L_i}},
% \end{equation}
% with the sum of all weights equal to 1. The width of a constructed prediction interval is the upper bound $U_i$ minus the lower bound $L_i$.

\section{Experiments and results}
\subsection{Battery Dataset}
The battery dataset used in the present work is originally from the work of Toyota Research Institute in collaboration with Stanford University and MIT \cite{severson2019data}. An early-prediction model developed in their work \cite{severson2019data} was later used for selecting high-cycle-life charging protocols \cite{attia2020closed}. There are 124 lithium iron ferrous phosphate (LFP)/graphite cells in this dataset with a nominal capacity of 1.1 Ah. The 124 cells are from 3 different test batches (i.e., the "2017-05-12" batch, the "2017-06-30" batch, and the "2018-04-12" batch) with batch date referring to the date the batch started. All the cells are tested at a constant temperature of 30 $^{\circ}$C in an environmental chamber. The cells are charged with a one-step or two-step fast-charging protocol and identically discharged at 4 C-rate. Cells are charged from 0\% to 80\% state-of-charge (SoC) with one of 72 charging protocols, for example, a charging protocol "5.6C(36\%)-4.3C" consists of a 5.6 C charging step from 0\% to 36\% SoC, followed by a 4.3 C step from 36\% to 80\% SoC.
% Definition of EOL in this work
It is assumed that the lithium-ion batteries reach their end of life (EOL) when their discharge capacity has decreased to 80\% of their initial nominal capacity.
% Data collection (V,I,T,R)
Time-series voltage, current, and cell temperature were continuously measured during cycling. The internal resistance was measured per cycle during charging at 80\% SoC by averaging 10 pulses of $\pm3.6$ C with a pulse width of around 30 ms.
\subsection{Train-test Split}
There are 72 different charging protocols in this battery dataset with nominal charging time from 0\% to 80\% SoC ranging from 9 to 13.3 minutes. For the purpose of reducing the possibly large sampling error due to the small dataset used in this work, the stratified random sampling method \cite{reitermanova2010data} is employed to randomly split the dataset into a training set that contains 80\% of the total dataset (99 samples) for optimizing model hyperparameters and learning the final model, and a test set that contains 20\% of the total dataset (25 samples) for evaluating the performance of the final model. At each split, equal ratios of fast-charged (i.e., less than 10.5 min) cells, medium-charged (i.e., between 10.5 and 11.7 min) cells, and slow-charged (i.e., greater than 11.7 min) cells are maintained in the training and test set. Moreover, in order to reduce the random effect of the selected split, stratified random sampling is repeated 5 times, and then the results of 5 train-test splits are averaged.

\subsection{Performance Evaluation and Results}
% Step 1 - the values of hyperparameters are reported.
For a fair comparison,  the optimal sets of hyperparameters for the QRF model and other benchmark models (i.e., Elastic Net regression model, GPR, and RVM) are obtained using the hyperparameter optimization method described in the previous section, given the same training set at each train-test split. The optimal sets of hyperparameters for the QRF model and the other models are reported in Table \ref{tab0}.
\begin{table*}[!htbp]
\caption{Optimal model hyperparameters}
\begin{center}
\begin{tabular}{|c|l|l|}
\hline
\textbf{Models} & \textbf{Model hyperparameters} & \shortstack{\textbf{Optimal values} \\ \textbf{5 train-test splits}}\\
\hline
Elastic net & \shortstack[l]{$\alpha$ - the relative weight of \\ the L1 and L2 penalties; \\ $\lambda$ - the regularization parameter.}  & \shortstack[l]{$\alpha_1=0.42$, $\lambda_1=0.0572$;  \\ $\alpha_2=1.00$, $\lambda_2=0.0313$; \\ $\alpha_3=0.01$, $\lambda_3=0.0218$; \\ $\alpha_4=0.01$, $\lambda_4=0.0333$; \\ $\alpha_5=0.57$, $\lambda_5=0.0945$.}\\
\hline
\shortstack{GPR \\ (Sum of a radial basis \\ function kernel and \\ a white noise kernel)} & \shortstack[l]{$\sigma_l$ - the length scale; \\ $\sigma^2_f$ - the signal variance; \\ $\sigma^2_n$ - the noise variance.} & \shortstack[l]{$\sigma_{l_1}=8.1393$, $\sigma^2_{f_1}=976.3646$, $\sigma^2_{n_1}=17508.5982$; \\ $\sigma_{l_2}=8.7730$, $\sigma^2_{f_2}=983.8729$, $\sigma^2_{n_1}=26238.2269$; \\ $\sigma_{l_3}=10.5000$, $\sigma^2_{f_3}=989.1063$, $\sigma^2_{n_3}=20682.1931$; \\ $\sigma_{l_4}=8.1573$, $\sigma^2_{f_4}=947.8936$, $\sigma^2_{n_4}=13712.8706$; \\ $\sigma_{l_5}=7.2395$, $\sigma^2_{f_5}=787.9068$, $\sigma^2_{n_5}=20715.8807$.}\\
\hline
\shortstack{RVM \\ (Radial basis \\ function kernel)} & \shortstack[l]{$\gamma$ - the kernel scale; \\ $\beta$ - the inverse noise variance.} & \shortstack[l]{$\gamma_1=0.01002$, $\beta_1=0.00003$; \\ $\gamma_2=0.01003$, $\beta_2=0.00003$; \\$\gamma_3=0.01009$, $\beta_3=0.00003$; \\$\gamma_4=0.01189$, $\beta_4=0.00004$; \\$\gamma_5=0.01397$, $\beta_5=0.00003$.} \\
\hline
QRF+AIS & \shortstack[l]{$n_{tree}$ - the number of trees; \\ $m_{try}$ - the number of random features \\ in each split; \\ $l_{node}$ - the minimum number of \\ samples at a leaf node.} & \shortstack[l]{$n_{{tree}_{1}}=213$, $m_{{try}_{1}}=9$, $l_{{node}_{1}}=1$; \\ $n_{{tree}_{2}}=1393$, $m_{{try}_{2}}=12$, $l_{{node}_{2}}=2$; \\ $n_{{tree}_{3}}=1776$, $m_{{try}_{3}}=12$, $l_{{node}_{3}}=3$; \\ $n_{{tree}_{4}}=110$, $m_{{try}_{4}}=6$, $l_{{node}_{4}}=1$; \\ $n_{{tree}_{5}}=191$, $m_{{try}_{5}}=11$, $l_{{node}_{5}}=4$.} \\
\hline
QRF+ALW & \shortstack[l]{$n_{tree}$ - the number of trees; \\ $m_{try}$ - the number of random features \\ in each split; \\ $l_{node}$ - the minimum number of \\ samples at a leaf node.} & \shortstack[l]{$n_{{tree}_{1}}=1184$, $m_{{try}_{1}}=8$, $l_{{node}_{1}}=4$; \\ $n_{{tree}_{2}}=982$, $m_{{try}_{2}}=8$, $l_{{node}_{2}}=6$; \\ $n_{{tree}_{3}}=1072$, $m_{{try}_{3}}=12$, $l_{{node}_{3}}=6$; \\ $n_{{tree}_{4}}=1753$, $m_{{try}_{4}}=8$, $l_{{node}_{4}}=1$; \\ $n_{{tree}_{5}}=854$, $m_{{try}_{5}}=10$, $l_{{node}_{5}}=4$.} \\
\hline
% \multicolumn{6}{l}{Sample of a Table footnote.}
\end{tabular}
\label{tab0}
\end{center}
\end{table*}

% Step 2 - the results of point and range prediction are reported
% Point prediction comparison results and discussion.
% \textcolor{red}{The point predictions, i.e., the predicted mean of cycle life as defined in Eqn. (\ref{eq: 20}) of the QRF model are evaluated by comparison with those of Elastic Net regression model originally used on this dataset \cite{severson2019data}, GPR, and RVM. Additionally, the QRF model is capable of providing range predictions, e.g., 95\% PI of cycle life as defined in Eqn. (\ref{eq: 11.1}). The range prediction performance of the QRF model using proposed ALW defined in Eqn. (\ref{eq: 23}) as PI evaluation criterion in LOO-XVE is evaluated and compared with that using AIS defined in Eqn. (\ref{eq: 12}).
% }
The final QRF model and other models with their optimal sets of hyperparameters are learned on the same training set, and then evaluated using the same test set at each train-test split, with evaluation metrics for point predictions or range prediction. The results of 5 train-test splits are averaged and reported in Table \ref{tab1} and Table \ref{tab2}, respectively.

To compare the point prediction and range prediction results between the QRF model and other models, the performance improvement in percentage is calculated with Elastic Net model as benchmark in the point prediction comparison, and with QRF using AIS defined in Eqn. (\ref{eq: 12}) as PI evaluation criterion in LOO-XVE as benchmark in the range prediction comparison.
% Reporting results, i.e., RMSE, MAPE, R^2, PICP, MPIW, AIS, performance improvement.
In terms of point prediction, it can be seen from the results in Table \ref{tab1} that the QRF outperforms Elastic Net regression model, GPR, and RVM, evaluated by all three performance measures, i.e., RMSE, MAPE and $R^2$. In terms of range prediction, it can be seen from the results in Table \ref{tab2} that even though the QRF model using the proposed ALW as PI evaluation criterion in LOO-XVE has a 11.2\% worse MPIW value than the QRF model using the AIS as PI evaluation criterion in LOO-XVE, it has a 4.9\% better PICP value that is much closer to the nominal coverage probability 95\%, and its overall evaluation of PIs via AIS is 10.3\% better than those by the QRF model using AIS. Notably, the GPR model has the best PICP value over all other models, but is not able to compete with the QRF model measured by MPIW or AIS.

% Summarizing the results and discussion
In summary, these results suggest that the QRF model is capable of providing at least 20\% higher point prediction accuracy than Elastic Net model whose effectiveness was demonstrated in the work of Severson et al. \cite{severson2019data}. A possible explanation would be that, as a non-parametric model, the QRF model is more flexible than parametric models (e.g., the Elastic Net model) to extract complex patterns in the battery data without necessarily incurring severe overfitting. Moreover, by using the proposed ALW as PI evaluation criterion in LOO-XVE, the range prediction performance by the QRF models has indeed improved: higher coverage probability that is closer to the preassigned nominal 95\% coverage probability guarantees higher reliability of the final QRF model. Even though GPR provides the best coverage probability, the QRF model outperforms GPR in terms of width of PIs evaluated by MPIW and comprehensive measure of both coverage probability and width of PIs by AIS. The final QRF model using the proposed ALW as PI evaluation criterion in LOO-XVE is the one used for later analyses.

\begin{table*}[!htbp]
\caption{Battery cycle life point prediction performance}
\begin{center}
\begin{tabular}{|c|c|c|c|c|c|c|}
\hline
\textbf{Models} & \multicolumn{3}{|c|}{\shortstack{\textbf{Point prediction} \\ \textbf{evaluation}}} & \multicolumn{3}{|c|}{\shortstack{\textbf{Performance improvement} (\%)}} \\
\cline{2-7} 
& RMSE (cycles) & MAPE (\%) & $R^2$ & RMSE (cycles) & MAPE (\%) & $R^2$ \\
\hline
Elastic net & 196 & 20.2 & 0.70 & / & / & / \\
\hline
GPR & 216 & 14.0 & 0.63 & -10.2\% & -30.6\% & -10.0\% \\
\hline
RVM & 226 & 16.0 & 0.60 & -15.3\% & -20.8\% & -14.3\% \\
\hline
QRF+AIS & 142 & 10.9 & 0.85 & 27.6\% & 46.0\% & 20.0\% \\
\hline
QRF+ALW & 158 & 12.0 & 0.81 & 19.4\% & 40.6\% & 15.7\% \\
\hline
% \multicolumn{6}{l}{Sample of a Table footnote.}
\end{tabular}
\label{tab1}
\end{center}
\end{table*}

\begin{table*}[!htbp]
\caption{Battery cycle life range prediction performance}
\begin{center}
\begin{tabular}{|c|c|c|c|c|c|c|}
\hline
\textbf{Models} & \multicolumn{3}{|c|}{\shortstack{\textbf{Range prediction} \\ \textbf{evaluation}}} & \multicolumn{3}{|c|}{\shortstack{\textbf{Performance improvement} (\%)}} \\
\cline{2-7} 
& PICP (\%) & MPIW (cycles) & AIS (cycles) & PICP (\%) & MPIW (cycles) & AIS (cycles) \\
\hline
GPR & 96.0 & 677 & 984 & 6.7\% & -54.6\% & -50.9 \\
\hline
RVM & 94.4 & 731 & 1602 & 4.9\% & -66.9\% & -145.7\% \\
\hline
QRF+AIS & 90.0 & 438 & 652 & / & / & / \\
\hline
QRF+ALW & 94.4 & 487 & 585 & 4.9\% & -11.2\% & 10.3\% \\
\hline
% \multicolumn{6}{l}{Sample of a Table footnote.}
\end{tabular}
\label{tab2}
\end{center}
\end{table*}

\subsection{Correlation Analysis and Results}
To statistically interpret the width of battery cycle life range prediction, hypothesis tests using Pearson's correlation coefficient are conducted. More specifically, we determine whether there is a linear correlation between width of range prediction and variables under investigation. The first step is to define the null hypothesis (i.e., $H_0$ - width of range prediction does not linearly correlate with the variable under investigation) and the alternative hypothesis (i.e., $H_1$ - width of range prediction linearly correlates with the variable under investigation). A $p$-value less than a significance level of 0.05 allows to reject the null hypothesis, which indicates that width of range prediction highly correlates with the variable under investigation \cite{rice2006mathematical}.

The first hypothesis test is performed to decide whether there is a linear correlation between width of range prediction and absolute mean prediction error (i.e., the distance between the point prediction and the observation) on the test set over 5 train-test splits. The second hypothesis test is performed to decide whether there is a linear correlation between width of range prediction and each of the 12 input features on the test set in the 5 train-test splits.

For the first hypothesis test, the Pearson's correlation coefficient is found to be 0.71 with a sufficiently low $p$-value that allows to reject the null hypothesis ($p$-value $<$ 0.05). We, therefore, conclude that there is sufficient evidence that width of range prediction and absolute mean prediction error are highly correlated.

For the second hypothesis test, the Pearson's correlation coefficient values and corresponding $p$-values between width of range prediction and the 12 input features are listed in Table \ref{tab3}, where the results are sorted in ascending order of their $p$-values. It can be seen from the results in Table \ref{tab3} that in total 6 input features are highly correlated with width of range prediction, and that the variance and minimum of difference of the discharge voltage curve between cycle 100 and cycle 10 (i.e., $\Delta Q_{100-10} (V)$) have the largest effect on the predicted range.

\begin{table*}[!htbp]
\caption{Hypothesis test results of correlation between width of range prediction and input features}
\begin{center}
\begin{tabular}{|c|c|c|c|}
\hline
\textbf{Input features} & \shortstack{\textbf{Pearson's correlation} \\ \textbf{coefficient values}} & \textbf{$p$-values} & \textbf{Decisions}\\
\hline
var\_dQ\_100\_10 & -0.563 & 1e-11 & rejected $H_0$ \\
\hline
minimum\_dQ\_100\_10 & -0.545 & 1e-10 & rejected $H_0$ \\
\hline
IR\_2 & -0.279 & 0.0016 & rejected $H_0$ \\
\hline
variance\_dT\_100\_10 & -0.275 & 0.0019 & rejected $H_0$ \\
\hline
intercept\_lin\_fit\_2\_100 & -0.239 & 0.0072 & rejected $H_0$\\
\hline
maximum\_IR\_2\_100 & -0.215 & 0.0160 & rejected $H_0$\\
\hline
peak\_amplitude\_dQdV\_100\_10 & -0.172 & 0.0551 & Failed to reject $H_0$\\
\hline
diff\_maximum\_T\_100\_10 & -0.166 & 0.0643 & Failed to reject $H_0$\\
\hline
peak\_position\_dQdV\_100\_10 & -0.155 & 0.0848 & Failed to reject $H_0$\\
\hline
diff\_mean\_T\_100\_10 & -0.136 & 0.1299 & Failed to reject $H_0$\\
\hline
minimum\_T\_10 & 0.129 & 0.1521 & Failed to reject $H_0$\\
\hline
diff\_IR\_100\_2 & -0.008 & 0.9255 & Failed to reject $H_0$\\
\hline
% \multicolumn{6}{l}{Sample of a Table footnote.}
\end{tabular}
\label{tab3}
\end{center}
\end{table*}

\subsection{Computational Aspects}
The whole experiment in this work runs on a laptop with an intel Core i5 CPU and 16 GB memory. The hyperparameter optimization for the QRF model during the training stage takes approximately 10 hours at one train-test split. Then the final QRF model with the optimal set of hyperparameters is learned on the training set. At the inference stage, it takes the final QRF model approximately 150 ms to predict the cycle life of a battery cell given its input feature realization from its early degradation data.

For the real application of the final QRF model, we must first investigate the viability of embedding the final QRF model on a BMS, for example, memory requirement of a final QRF model implementation on a BMS. If the final QRF model is heavy then we will prefer to implement a large part of it in the cloud whereas preprocessing of data and feature engineering can be done on-board so that we do not have to send high frequency data to the cloud via vehicle's telemetry gateway. Moreover, the main purpose of this model is to perform early prediction of cycle life. Since aging is a slow process, the final QRF model does not need to run in real-time. We may run it periodically or in an event-triggered manner in the cloud based on preprocessed data from on-board BMS.

\section{Model interpretation}
\begin{figure*}[!ht]
\centering
\subfloat[Permutation importance ranking on the training set]{\includegraphics[width=2.5in]{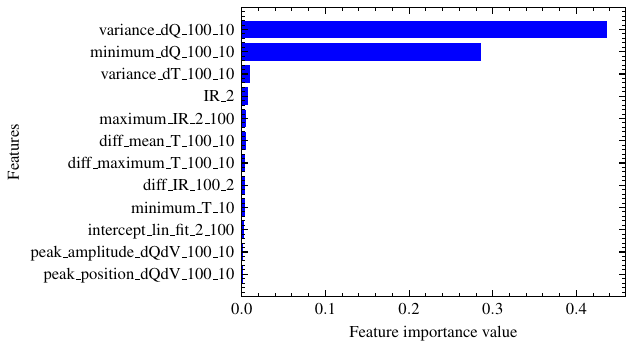}%
\label{fig2_first_case}}
\hfil
\subfloat[Permutation importance ranking on the test set]{\includegraphics[width=2.5in]{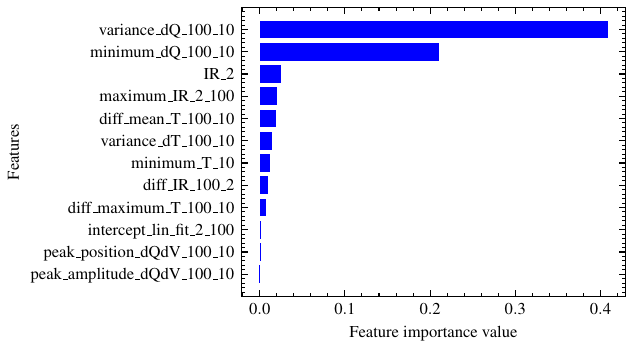}%
\label{fig2_second_case}}
\caption{Permutation importance ranking of 12 selected features.}
\label{fig_2}
\end{figure*}

Permutation importance is computed on the training set as shown in Fig. \ref{fig2_first_case}. The two most important features identified are related to the $\Delta Q_{100-10} (V)$ curve, i.e., the discharge capacity change as a function of voltage between cycle 100 and cycle 10, which was selected in the cycle life prediction model developed by Severson et al. \cite{severson2019data}. This $\Delta Q(V)$ curve is of great interest because the curve itself and its derivatives contain rich information for degradation diagnosis \cite{bloom2010differential} \cite{birkl2017degradation} \cite{berecibar2016online}. Based on the training set, the most important feature is the variance of $\Delta Q_{100-10} (V)$ curve (see \cite{severson2019data} for the definition of this feature), which means that the QRF model relies on this feature the most for making predictions. However, measuring the feature importance on the training set on which the QRF model is trained is not as informative as that on the unseen data. If the QRF model is overfitted, the feature importance measured on the training set may mislead us to believe that the wrong features are important. Therefore, the feature importance is also measured on the test set. It is shown in Fig. \ref{fig2_second_case} that the most importance feature is still the variance of $\Delta Q_{100-10} (V)$ curve, which means that this feature does indeed contribute the most to the prediction performance of the QRF model.

% \FloatBarrier

\begin{figure*}[!ht]
\centering
\subfloat[Variance of $\Delta Q_{100-10} (V)$]{\includegraphics[width=2.5in]{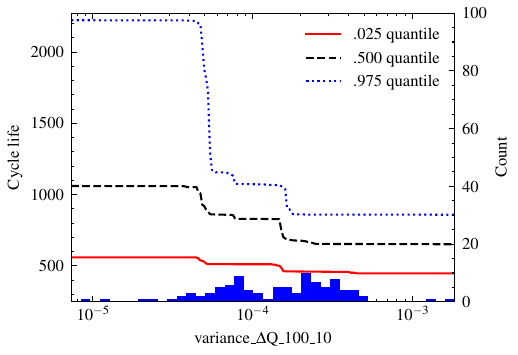}%
\label{fig3_first_case}}
\hfil
\subfloat[Minimum of $\Delta Q_{100-10} (V)$]{\includegraphics[width=2.5in]{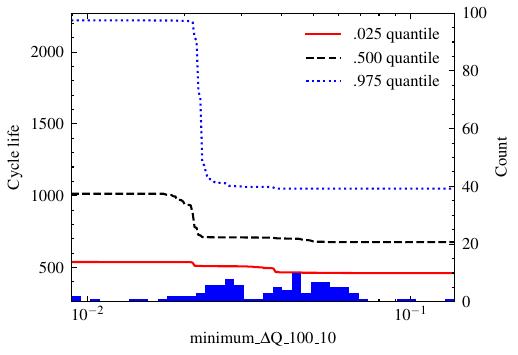}%
\label{fig3_second_case}}
\caption{PDPs for the cycle life prediction w.r.t. variance of $\Delta Q_{100-10} (V)$ and minimum of $\Delta Q_{100-10} (V)$. Histogram at the bottom shows observations of the feature, with scale to the right.}
\label{fig_3}
\end{figure*}

In order to further illustrate how the most important feature affects the predicted cycle life, a one-dimensional partial dependence plot (PDP) is computed on the training set (see Fig. \ref{fig_3}). A lower bound of the predicted battery cycle life as a function of variance of $\Delta Q_{100-10} (V)$ is provided by the .025 quantile curves, while the .975 quantile curve provides an upper bound of the predicted battery cycle life as a function of variance of $\Delta Q_{100-10} (V)$. The median value of the predicted battery cycle life is provided by the .50 quantile curves. The histogram on the x-axis shows the distribution of the observations of the variance of $\Delta Q_{100-10} (V)$ in the training set. The quantile curves flatten out in the regions of sparse observations of the variance of $\Delta Q_{100-10} (V)$ in the training data, thus not providing much information. In the region of dense distribution of variance of $\Delta Q_{100-10} (V)$ in training set, battery cycle life rapidly decreases when the variance of $\Delta Q_{100-10} (V)$ increases from $10^{-5}$ to $10^{-3}$, which indicates that a small increase of variance of $\Delta Q_{100-10} (V)$ during discharge has a large effect on battery degradation rate.
% Physical meaning of var_dQ_100_10
The physical meaning of the variance of $\Delta Q_{100-10} (V)$ is associated with the dependence of discharge energy dissipation on voltage. The variance of $\Delta Q_{100-10} (V)$ reflects the degree of non-uniformity in the discharge energy dissipation with voltage  \cite{severson2019data}. Thus, the larger the value of the variance of $\Delta Q_{100-10} (V)$ the larger the degree of non-uniformity in the discharge energy dissipation under galvanostatic conditions, which is consistent with the monotonic relationship between the variance of $\Delta Q_{100-10} (V)$ and cycle life (Fig. \ref{fig3_first_case}). The second most importance feature is the minimum of $\Delta Q_{100-10} (V)$, for which battery cycle life decreases with increasing value of the minimum of $\Delta Q_{100-10} (V)$, but not to the same extent as for the variance of $\Delta Q_{100-10} (V)$.

% \FloatBarrier

% Start discussion on predictive uncertainty
The PIs to the left in Fig. \ref{fig3_first_case} and on both ends (i.e., left and right) of the x-axis in Fig. \ref{fig3_second_case} are larger than those in the middle of the x-axis. The reason is that observations are lacking on both ends of the x-axis, and therefore, the learned QRF model is not confident to make predictions on these two zones, which leads to larger prediction uncertainty represented by the width of PIs. 

Furthermore, we would like to point out that data uncertainties are not considered in the present work, and it is assumed that all measurements are accurate and taken as if they were true.

The PDP results further illustrate how the variance of $\Delta Q_{100-10} (V)$ and the minimum of $\Delta Q_{100-10} (V)$ affect the predictions of battery cycle life quantitatively. Similarly, PDPs can be computed for all other features used as inputs to the QRF model on the training set.

% Why features extracted from high-rate discharge voltage curve have high predictive power and which degradation mode(s) cause this discharge voltage shift?
Severson et al. \cite{severson2019data} rationalized highly predictive features extracted from early-cycle discharge voltage curves (i.e., the variance of $\Delta Q_{100-10} (V)$ and the minimum of $\Delta Q_{100-10} (V)$) by experimentally investigating degradation modes that do not lead to immediate capacity fade but are manifested in the discharge voltage curves. They found out that loss of active material of the delithiated negative electrode contributes to a shift in the discharge voltage curve, with no change in capacity fade at early cycles. At high number of cycles, loss of active material of the delithiated negative electrode induces lithium plating, which irreversibly accelerates capacity loss. This degradation behavior is consistent with the high feature importance of variance and minimum of $\Delta Q(V)$, as shown in Fig. \ref{fig_2}. Throughout the literature, this degradation behavior is widely observed at various ambient temperature (e.g., 23 \textcelsius \ \cite{ansean2016fast} \cite{ansean2017operando}, 30 \textcelsius \ \cite{sarasketa2015understanding}, and 45 \textcelsius \ \cite{safari2011aging}) when the negative electrode capacity is larger than that of the positive electrode, as in case of LFP cells that we used in this work. Therefore, if these two features, (i.e., minimum and variance of $\Delta Q_{100-10} (V)$) are extracted from early degradation data under different ambient temperatures other than 30 \textcelsius, then the resulting learned QRF model based on this training dataset may still provide accurate cycle life point and range prediction.

\section{Application cases}
\begin{figure}[!ht]
\centering
\includegraphics[width=2.5in]{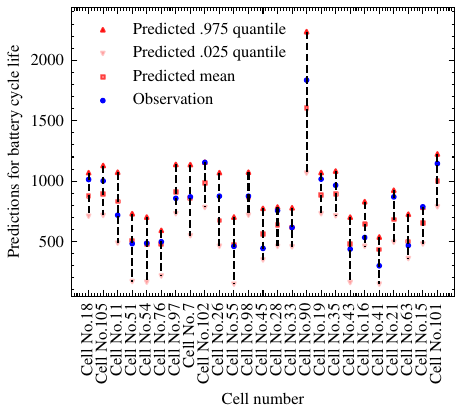}
\caption{The 95\% PIs and mean predictions by QRF.}
\label{fig_7}
\end{figure}

Based on predicted quantiles, PIs of battery cycle life can be constructed. To examine the prediction performance of the final QRF model by using it in a more intuitive way, we show in Fig. \ref{fig_7} the 95\% PIs and the mean predictions made by QRF on the test set at one split. There are 24 out of 25 observed cycle life samples within the PIs resulting in 96\% PICP, close to nominal  95\% coverage probability. This indicates that the constructed PIs with 95\% coverage probability exhibits good coverage probability of the observed cycle life values, which is required for a reliable battery cycle life range prediction.

% Discussion on applications of range predictions - The lower bound and upper bound of PIs (regarded as information) would facilitate 
The upper bound of the 95\% PI is the 97.5\% quantile prediction of cycle life, which means that battery cycle life may exceed the upper bound with a probability of around 2.5\%. Correspondingly, the lower bound of the 95\% PI is the 2.5\% quantile prediction of battery cycle life, which means that battery cycle life may fall below the lower bound with a probability of around 2 .5\%. The lower bound of the PIs would facilitate conservative decision-making while the upper bound would facilitate optimistic decision-making on battery applications.

% df_qrf_pi_RFE_ALW_final.json random_seed = 1

To demonstrate how the learned QRF model facilitates decision-making when the expected battery cycle life that is calculated using (Eqn. \ref{eq: ecl1}) under two different charging protocols are very close to each other, two groups of cells are selected where each group contains three cells from the same batch date and charged with the same charging protocol (see Table \ref{tab4}).

\begin{figure*}[!ht]
\centering
\subfloat[Predicted mean and 95\% prediction interval of Cell No.109, No.117 and No.121]{\includegraphics[width=2.5in]{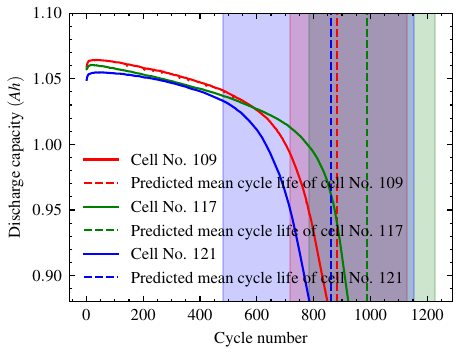}%
\label{fig9_first_pair}}
\hfil
\subfloat[Predicted mean and 95\% prediction interval of Cell No.94, No.96 and No.116]{\includegraphics[width=2.5in]{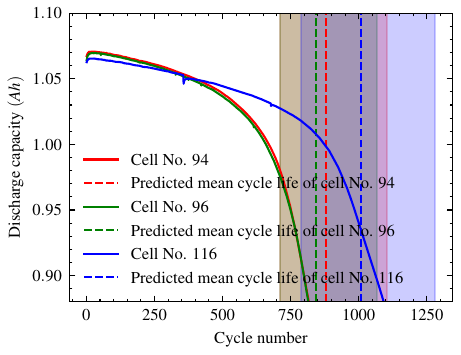}%
\label{fig9_second_pair}}
\caption{Selected two groups of cells.}
\label{fig_9}
\end{figure*}

% Step 1 - Showing the values of observed cycle life of two cells and calculating the true battery cycle life for each charging protocol
% Step 2 - Showing the values of predicted mean cycle life and width of range prediction and calculating the expected battery cycle life for each charging protocol.
% Step 3 - Selecting the high-cycle-life charging protocol out of these two protocols based on calculated expected battery cycle life for each charging protocol, discussing around the choice and motivating the advantages of the proposed EBCL2.
% Step 1 - Showing the values of observed cycle life of two cells and calculating the true battery cycle life for each charging protocol
The first group contains cell No.109, No.117 and No.121. All of them are from the batch date "2018-04-12" and charged with the charging protocol "5.6C(36\%)-4.3C". As one can see in Fig. \ref{fig9_first_pair}, three cells exhibit different capacity fade curves. The true battery cycle life of this charging protocol is calculated as the averaged observed cycle lives of three cells, i.e., 853 cycles. The second group contains cell No.94, No.96 and No.116. All of them are from the batch date "2018-04-12" and charged with the charging protocol "5.6C(19\%)-4.6C". As one can see in Fig. \ref{fig9_second_pair}, cell No.94 and No.96 exhibit very similar capacity degradation curve while the cell No.116 exhibits longer cycle life. The true battery cycle life of this charging protocol is calculated as the averaged observed cycle lives of three cells, i.e., 909 cycles. Therefore, based on the true cycle lives of two charging protocols, the charging protocol "5.6C(19\%)-4.6C" should be selected as the high-cycle-life charging protocol.
% Step 2 - Showing the values of predicted mean cycle life and width of range prediction and calculating the expected battery cycle life for each charging protocol.
% The predicted mean cycle lives of three cells in the first group are 849 cycles for cell No.91 and 1555 cycles for cell No.114. The width of range prediction varies significantly in this group of two cells with 369 cycles for cell No.91, and 1318 cycles for cell No.114, which suggests that the predicted mean cycle life of cell No.114 might deviate farther from its observed cycle life than that of cell No.91. 
The expected battery cycle life of the charging protocol "5.6C(36\%)-4.3C", calculated using Eqn. (\ref{eq: ecl1}), is equal to 911 cycles, while the expected battery cycle life of the charging protocol "5.6C(19\%)-4.6C", calculated using Eqn. (\ref{eq: ecl1}), is also equal to 911 cycles. 
% In the second group, the predicted mean cycle lives of two cells are 879 cycles for cell No.94 and 1001 cycles for cell No.101. The width of range prediction of cell No.94 is similar to that of cell No.101 (i.e., 392 cycles for cell No.94, 439 cycles for cell No.101), suggesting a possible similar discrepancy between predicted mean cycle life and the observed cycle life. 
% The expected battery cycle life 1 of the charging protocol "5.6C(19\%)-4.6C" calculated using Eqn. (\ref{eq: ecl1}) is equal to 940 cycles, while the expected battery cycle life 2 calculated using Eqn. (\ref{eq: ecl2}) is equal to 937 cycles. 
% Step 3 - Selecting the high-cycle-life charging protocol out of these two protocols based on calculated expected battery cycle life for each charging protocol, discussing around the choice and motivating the advantages of the proposed EBCL2.
With point prediction alone, it is very difficult to select the high-cycle-life charging protocol out of two charging protocols as they have the same expected battery cycle lives. However, with additional range prediction, the expected battery cycle life range of each charging protocol can be calculated using Eqn. (\ref{eq: ecl2}), i.e., 434 cycles for the charging protocol "5.6C(36\%)-4.3C" and 392 cycles for the charging protocol "5.6C(19\%)-4.6C". Therefore, the second charging protocol (i.e., ”5.6C(19\%)-4.6C”) is chosen here as the preferred among two due to lower uncertainty (and hence lower economic risk) around the expected battery cycle life.

\begin{table*}[!htbp]
\caption{Two groups of cells with two charging protocols respectively}
\begin{center}
\begin{tabular}{|l|c|c|c|c|c|c|}
\hline
\multicolumn{1}{|l|}{\textbf{Batch date}} & \multicolumn{6}{|c|}{\textbf{2018-04-12}}\\
\hline
 \multicolumn{1}{|l|}{\textbf{Charging protocols}} & \multicolumn{3}{|c|}{\shortstack{\textbf{5.6C(36\%)-4.3C}}} & \multicolumn{3}{|c|}{\shortstack{\textbf{5.6C(19\%)-4.6C}}} \\
\hline
\multicolumn{1}{|l|}{\textbf{Cell number}} & \textbf{No.109} & \textbf{No.117} & \textbf{No.121} & \textbf{No.94} & \textbf{No.96} & \textbf{No.116}\\
\hline
\shortstack[l]{Observed cycle life (cycles)} & 850 & 923 & 786 & 817 & 816 & 1093 \\
\hline
\shortstack[l]{Predicted mean cycle life (cycles)} & 884 & 989 & 860 & 879& 845& 1008 \\
\hline
\shortstack[l]{The lower bound of range prediction (cycles)} & 718 & 785 & 481 & 712 & 711 & 787 \\
\hline
\shortstack[l]{The upper bound of range prediction (cycles)} & 1128 & 1229 & 1152 & 1104 & 1069 & 1282 \\
\hline
\shortstack[l]{True battery cycle life of \\ a charging protocol (cycles)} & \multicolumn{3}{|c|}{853} & \multicolumn{3}{|c|}{909}\\
\hline
\shortstack[l]{Expected battery cycle life of \\ a charging protocol (cycles)} & \multicolumn{3}{|c|}{911} & \multicolumn{3}{|c|}{911}\\
\hline
\shortstack[l]{Expected battery cycle life range of \\ a charging protocol (cycles)} & \multicolumn{3}{|c|}{434} & \multicolumn{3}{|c|}{392}\\
\hline
% \multicolumn{6}{l}{Sample of a Table footnote.}
\end{tabular}
\label{tab4}
\end{center}
\end{table*}

In the two examples given above, it can be seen how the discrepancy between predicted mean cycle life and the observed cycle life of a cell agrees with the width of range prediction. The widths of range prediction may provide more information for decision-making (for instance, when selecting high-cycle-life charging protocol) under uncertainty associated with cycle life prediction than we get from single point predictions alone. Interestingly, the differences in observed cycle life within each group can be quite remarkable (see Fig. \ref{fig9_first_pair} and Fig. \ref{fig9_second_pair}), considering that they are expected to have almost the same aging process. The reason why cells from the same batch and charged with the same charging protocols can have a very different cycle life is most likely due to production-related factors (e.g., the variance of material properties and process parameters). Apparently, in this case, the cell-to-cell variations caused by the production process are quite significant, and these variations are indeed captured by the model via the extracted features.

% For the same batch of Li-ion battery cells, it is particularly interesting with cells having approximately the same point predictions but different range of cycle life, since that gives an indication that there are differences in their operating conditions that has a large effect on their cycle life assuming they are identical at start. For example, when selecting optimal charging protocol, further investigation should be taken if two identical cells charged with two different charging policies have the close cycle life point prediction but different cycle life range.

%  Firstly, we consider the pair consisting of cell No.16 and cell No.60. Both of them have very close mean prediction of cycle life (608 cycles for cell No. 16 and 604 cycles for cell No.60 respectively). However, cell No.60 has a wider prediction interval compared to cell No.16, although the prediction interval successfully contains the observed cycle life of cell No.60. The mean prediction error of cell No.60 is larger than that of cell No.16. After plotting capacity degradation curve for both two cells in Fig. \ref{fig9_first_pair}, it can be seen that cell No.60 degrades at low rate before 300 cycles followed by a sudden accelerated degradation. In fact, cell No.16 is from the "2017-05-12" batch with charging protocol "5.4C(60\%)-3.6C", while cell No.60 is from the "2017-06-30" batch with charging protocol "4.8C(80\%)-4.8C". In order to 

\section{Conclusion}
In the present paper, we have proposed a quantile regression forest (QRF) model for Li-ion battery cycle life prediction using early degradation data. To the best of our knowledge, it is the first time that the QRF model is introduced to battery cycle life range prediction even though it has been used in predictions of drug effect, crop yield etc. The proposed PI evaluation criterion in LOO-XVE can be used for optimizing hyperparameters of other regression models that are capable of providing range predictions so that high prediction interval coverage probability (PICP) value, which satisfies a preassigned nominal coverage probability, as well as low mean prediction interval width (MPIW) value can be available for decision-making. Two global model-agnostic methods were employed to interpret the final QRF model, and they can be easily employed also for other advanced data-driven methods. These interpretation techniques can reveal underlying battery aging mechanisms and help finding features that have the highest predictive power for cycle life prediction with data-driven methods. There are, however, several improvements that can be made in future work. First, the battery dataset used for training and testing is relatively small with cells tested at ambient temperature of 30 \textcelsius. A larger battery dataset with cells tested at ambient temperatures other than 30 \textcelsius \ is desired for both validating prediction performance of the proposed QRF model and effectiveness of the proposed two interpretation techniques that may reveal the underlying degradation process of cells tested at other ambient temperatures. Second, permutation importance and PDP are the two methods used in the present work to implicitly interpret the model. Additional interpretation techniques can be introduced in the future for further interpretation of other advanced data-driven models. Thirdly, towards online application of the final QRF model on a realistic BMS, several aspects need to be investigated, including, for example, computational efficiency and memory footprint of the final QRF model for real-time embedded applications. Fourthly, it would be very interesting to investigate the robustness of the QRF model with respect to both varying operating conditions and measurement noise in the data. Fifthly, our proposed method does not consider calendar aging that happens during any dedicated resting periods for long time (e.g., battery storage, vehicle parking etc). Since the calendar aging impacts the early capacity fade of a cell, it is reasonable to expect that a QRF model trained using calendric and cyclic aging data may still show good prediction performance. However, this needs to be tested and verified in our future work. Lastly, the hybrid data-driven method is also an important category, one possibility of using a second data-driven model to extrapolate values of important features can be investigated so that even earlier or less degradation data is needed for battery cycle life prediction with high accuracy and reliability. Another possibility of using physics-based model for plausibility check and rationalization of prediction output from data-driven model can also be investigated. This may help interpret confidence in the predictions by the data-driven model.

% if have a single appendix:
%\appendix[Proof of the Zonklar Equations]
% or
%\appendix  % for no appendix heading
% do not use \section anymore after \appendix, only \section*
% is possibly needed

% use appendices with more than one appendix
% then use \section to start each appendix
% you must declare a \section before using any
% \subsection or using \label (\appendices by itself
% starts a section numbered zero.)
%

%\appendices
%\section{Proof of the First Zonklar Equation}
%Appendix one text goes here.

% you can choose not to have a title for an appendix
% if you want by leaving the argument blank
%\section{}
%Appendix two text goes here.

% use section* for acknowledgment
\section*{Acknowledgment}
The authors would like to thank Volvo AB and Swedish Energy Agency for funding this work with project 45540-1. In particular, the authors would like to thank Xixi Liu from the Department of Electrical Engineering, Chalmers University of Technology for her constructive suggestions on the structure of the manuscript, and Che-Tsung Lin from the Department of Electrical Engineering, Chalmers University of Technology for the discussion on hyperparameter optimization.

% Can use something like this to put references on a page
% by themselves when using endfloat and the captionsoff option.
\ifCLASSOPTIONcaptionsoff
  \newpage
\fi

% trigger a \newpage just before the given reference
% number - used to balance the columns on the last page
% adjust value as needed - may need to be readjusted if
% the document is modified later
%\IEEEtriggeratref{8}
% The "triggered" command can be changed if desired:
%\IEEEtriggercmd{\enlargethispage{-5in}}

% references section
\bibliographystyle{IEEEtran}
\bibliography{References}
\end{document}